\documentclass[preprint]{revtex4-1}
\usepackage{graphicx}
\usepackage{bm}
\usepackage{amsmath,amsfonts,amssymb}
\usepackage[utf8]{inputenc}
\newcommand {\be} {\begin{equation}}
\newcommand {\ba} {\begin{eqnarray}}
\newcommand {\ee} {\end{equation}}
\newcommand {\ea} {\end{eqnarray}}

\begin{document}

\title{Pomeron and Reggeon contributions to elastic proton-proton and proton-antiproton scattering in holographic QCD}

\author{Zhibo Liu}
\email{202107020021014@ctgu.edu.cn}
\affiliation{College of Science, China Three Gorges University, Yichang 443002, People's Republic of China}
\affiliation{Center for Astronomy and Space Sciences, China Three Gorges University, Yichang 443002, People's Republic of China}
\author{Wei Xie}
\email{xiewei@ctgu.edu.cn}
\affiliation{College of Science, China Three Gorges University, Yichang 443002, People's Republic of China}
\affiliation{Center for Astronomy and Space Sciences, China Three Gorges University, Yichang 443002, People's Republic of China}
\author{Akira Watanabe}
\email{watanabe@ctgu.edu.cn (Corresponding author)}
\affiliation{College of Science, China Three Gorges University, Yichang 443002, People's Republic of China}
\affiliation{Center for Astronomy and Space Sciences, China Three Gorges University, Yichang 443002, People's Republic of China}

\date{\today}

\begin{abstract}
The total and differential cross sections of elastic proton-proton and proton-antiproton scattering are studied in a holographic QCD model, considering the Pomeron and Reggeon exchanges in the Regge regime.
In our model setup, the Pomeron and Reggeon exchanges are described by the Reggeized spin-2 glueball and vector meson propagators, respectively.
How those contributions change with the energy is explicitly shown, focusing on the contribution ratios.
The adjustable parameters included in the model are determined with the experimental data, and it is presented that the resulting total and differential cross sections are consistent with the data in a wide kinematic region.
\end{abstract}

\maketitle

\section{Introduction}
\label{sec:introduction}
Quantum chromodynamics (QCD), being a remarkably successful theory of the strong interaction, meets some practical difficulties in the soft kinematic region.
One of those difficulties is related to the hadron-hadron forward scattering processes~\cite{Fiore:2008tp,Jenkovszky:2018itd}, with high center-of-mass energy $s$ and small momentum transfer $t$.
Decades ago, the scaling laws for the hadron-hadron cross sections at $s \to \infty$ and fixed $t/s$ were investigated~\cite{Brodsky:1973kr,Matveev:1973ra,Lepage:1980fj}, but the story is totally different for the forward scattering due to the nonperturbative nature of the involved partonic dynamics.
The forward scattering is associated with the hadron diffraction, in which the initial and final states in the scattering process have the same quantum numbers.
The diffractive processes are generally classified as double diffraction dissociation, single diffraction dissociation and elastic scattering.
The simplest process is elastic hadron-hadron scattering, in which the imaginary part of the scattering amplitude is related to the total cross 
section.
Thus, the investigation of total cross section is embedded into the elastic scattering amplitude.
The total cross sections of various hadronic processes have been measured experimentally~\cite{ParticleDataGroup:2020ssz} and analyzed phenomenologically~\cite{Cudell:1999tx,Cudell:2001pn,COMPETE:2002jcr,Cudell:2019mbe}.
These measurements provide crucial information for the soft kinematic region of QCD, in which quarks and gluons are strongly interacting and perturbative method is basically not applicable.

Historically the Regge theory provided a useful framework to analyze the total cross sections.
Even today, the Regge theory, with combined Reggeon and Pomeron contributions, unrelated to the underlying partonic dynamics, still provides successful descriptions of total cross sections of hadronic scattering.
The Regge theory is based on the analysis with the complex angular momentum, which is extensively illustrated in the literature~\cite{Collins:1971ff,Collins:1977jy}.
The total cross sections of proton-proton ($pp$) and proton-antiproton ($p \bar p$) scattering have been fitted by the exchanges of the Reggeon trajectories and the soft Pomeron~\cite{Collins:1973uad,Collins:1974en,Donnachie:1992ny,Donnachie:2019ciz}, with their respective slope and intercept parameters.
The nature of the soft Pomeron is highly nonperturbative, and practically it is almost impossible to be derived from QCD.
Its properties can be inferred from experimental data accumulated over several decades.
The $2^{++}$ glueball is considered to be the lightest state on the leading Pomeron trajectory, which has an intercept of 1.08.
The increasing behavior of the total cross sections with respect to the center-of-mass energy $\sqrt s$ is associated with the Pomeron exchange.
On the other hand, the exchange of the Reggeon trajectories accounts for the decreasing behavior.

Holographic QCD, a nonperturbative approach to QCD, has been developed based on the anti-de Sitter/conformal field theory (AdS/CFT) correspondence~\cite{Maldacena:1997re,Gubser:1998bc,Witten:1998qj}, which, also called gauge-string duality, relates a strongly coupled gauge theory to a weakly coupled string theory in the curved spacetime.
Most works in holographic QCD assume that the curved space string theory reduces to the supergravity theory in vanishing string length.
This approach has been employed to analyze the spectrum and structure of hadrons with successful results~\cite{Erlich:2005qh,deTeramond:2005su,Karch:2006pv,Brodsky:2007hb,Abidin:2008ku,Abidin:2008hn,Abidin:2009hr,Branz:2010ub,Gutsche:2011vb,Li:2013oda,Brodsky:2014yha,Gutsche:2017lyu,Lyubovitskij:2020gjz}.
Holographic QCD has also been used to study high energy scattering processes~\cite{Polchinski:2001tt,Polchinski:2002jw,Brower:2006ea,Hatta:2007he,Pire:2008zf,Marquet:2010sf,Watanabe:2012uc,Watanabe:2013spa,Watanabe:2015mia,Watanabe:2018owy,Xie:2019soz,Burikham:2019zbo,Watanabe:2019zny,Liu:2022out}.
In particular, hadronic scattering in the Regge regime, which is characterized by the condition, $s\gg t$, is interesting.
Historically string theory originated in the phenomenological description of hadronic scattering in the Regge regime, in which the scattering amplitude is accounted by the exchange of the Regge trajectories of mesons.

Studies of high energy scattering in the Regge regime have gathered interests in the high energy physics community so far, since this regime is essential to better understanding and testing the gauge-string duality.
The strict treatment of the string dual requires complicated string calculations in the curved space.
Approximations are certainly needed to make practical progress in this field.
A string theory inspired holographic QCD model for hadronic scattering~\cite{Domokos:2009hm,Domokos:2010ma,Anderson:2016zon}, which aims to describe the experimental data of scattering cross sections, has been developed.
In this model the string scattering amplitude in the weakly curved background approximately takes the same structure as the flat space amplitude, with the Regge parameters (Regge slopes and intercepts) which differ from those in the flat space.
The amplitude in the Regge regime is dictated by the amplitude of exchange of lightest states on the Regge trajectories of mesons and glueballs, which are determined by low energy effective couplings in the top-down holographic approach~\cite{Sakai:2004cn,Iatrakis:2016rvj,Anderson:2016zon}.
The top-down construction contains relatively fewer parameters and unambiguously gives the hadron couplings via the supergravity action.
The couplings involving the proton, mesons and glueballs have been derived explicitly in the literature~\cite{Sakai:2004cn,Hong:2007ay,
Hong:2007dq,Domokos:2009cq,Anderson:2016zon}.
Based on these couplings, hadronic scattering amplitudes in the Regge regime can be obtained for exchanges of the lightest mesons or glueballs.
Then the single particle propagators are replaced with the Reggeized ones which are obtained by comparing with the string scattering amplitudes~\cite{Domokos:2009hm,Anderson:2016zon}.

This model has been applied to the proton-proton scattering via the Pomeron exchange~\cite{Domokos:2009hm,Domokos:2010ma}, and to the central production of $\eta$ with the double Pomeron and double Reggeon exchange~\cite{Anderson:2016zon}.
The propagator for proton-proton scattering via the Pomeron exchange is based on a comparison of the Virasoro-Shapiro amplitude with the classical flat space bosonic closed string amplitude.
The mesonic Regge trajectories have also been incorporated into this model, whose propagators are constructed by using the bosonic open string amplitude in flat space.
The parameters of these amplitudes are modified to account for the physical meson and glueball trajectories.

In this work we study the elastic $pp$ and $p \bar p$ scattering in the Regge regime, taking into account both the Pomeron and Reggeon exchanges.
This work is an extension of the preceding study~\cite{Xie:2019soz}, in which only the Pomeron exchange was considered.
Since the Pomeron exchange gives the dominant contribution to cross sections in the high energy region with $\sqrt{s} \gtrsim 100$~GeV, successful results were obtained in the previous work.
However, the Reggeon exchange contribution needs to also be considered to describe the data in the lower energy region.
In our model the Pomeron and Reggeon exchanges are described by the Reggeized spin-2 glueball and vector meson propagators, respectively.
The scattering amplitudes are obtained by combining the proton-vector meson and proton-glueball couplings with those propagators.

The model includes several parameters, but the ones related to the Pomeron exchange have already been determined in the previous work.
Hence, in this study we use the parameter values obtained in Ref.~\cite{Xie:2019soz} for the Pomeron, and determine the adjustable parameters associated with the Reggeon exchange by numerical fits with experimental data of the total cross sections for the $pp$ and $p \bar p$ scattering.
Utilizing the linear relation for the Reggeon trajectory, only three parameters in total need to be determined in this study.
With the obtained parameter values, we explicitly show how the both contributions change with the energy, focusing on the contribution ratios.
It is presented that the resulting total cross sections are in agreement with the data in a wide kinematic region.
Then, we show that our predictions for the differential cross section are also consistent with the data for both the $pp$ and $p \bar p$ scattering in the kinematic range of $|t| < 0.45$~GeV$^2$ and $10$~GeV $< \sqrt{s} \leq 13$~TeV.
Overall, the results presented in this paper show that the present model works well and the future experimental data could better constrain the model parameters.

This paper is organized as follows.
In Sec.~\ref{sec:model} we introduce the holographic description of the elastic $pp$ and $p \bar p$ scattering in the Regge regime, taking into account the Pomeron and Reggeon exchanges.
We briefly review the formalism developed in the preceding studies, and present the expressions for the total and differential cross sections.
The energy dependence of the both contributions is shown in detail, focusing on the contribution ratios.
With the obtained expressions, numerical fits are performed, and the results are displayed in Sec.~\ref{sec:results}.
Our conclusion with the implications of this work is given in Sec.~\ref{sec:conclusion}.

\section{Model setup}
\label{sec:model}

\subsection{Holographic description of the $pp$ and $p \bar p$ scattering}
\label{sec:subsec_formalism}
\begin{figure}[t]
\begin{center}
\begin{tabular}{ccccc}
\includegraphics[width=0.3\textwidth]{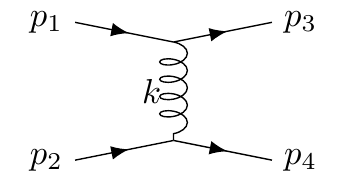}
\includegraphics[width=0.3\textwidth]{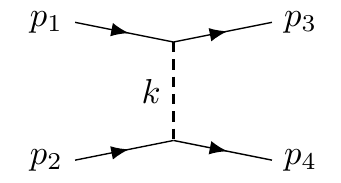}
\end{tabular}
\end{center}
\caption{
The left and right Feynman diagrams represent the $pp$($p\bar{p}$) scattering with the Pomeron and Reggeon exchanges in the $t$-channel, respectively.
$p_1, p_2$ and $p_3, p_4$ are four-momenta of the initial and final states, respectively, and $k$ is the momentum transfer.
}
\label{Feynman}
\end{figure}

In this work we investigate the elastic $pp$ and $p \bar p$ scattering in the Regge regime, considering contributions of the Pomeron and Reggeon exchanges which are described by the Reggeized spin-2 glueball and vector meson.
The Feynman diagrams, which describe those contributions, are shown in Fig.~\ref{Feynman}.
The scattering amplitudes are written as
\begin{equation}\label{tot}
      \mathcal{A}_{\rm tot}^{pp,p\bar{p}} = \mathcal{A}_{ g}^{pp,p\bar{p}} + \mathcal{A}_{ v}^{pp,p\bar{p}}.
\end{equation}
Following the preceding study~\cite{Pagels:1966zza}, the matrix element of the energy momentum tensor $T_{\mu\nu}$ between the initial and final proton states is expressed as
\begin{align}\label{em}
\langle p',s'|T_{\mu\nu}|p,s\rangle  =\bar{u}(p',s')\bigg[&A(t)\frac{\gamma_{\mu}P_{\nu}+\gamma_{\nu}P_{\mu}}{2}\nonumber
 \\&+B(t)\frac{i(P_{\mu}\sigma_{\nu\rho}+P_{\nu}\sigma_{\mu\rho})k^\rho}{4m_{p}}\nonumber
 \\&+C(t)\frac{k_{\mu}k_{\nu} - \eta_{\mu\nu}k^2}{m_{p}}\bigg]u(p,s),
\end{align}
where $P = (p_1 + p_3)/2 $, $k = p_3 - p_1$, $t=-k^{2}$ and $m_{p}$ is the proton mass.
$A(t)$, $B(t)$ and $C(t)$ are the proton form factors.
The contributions of $B(t)$ and $C(t)$ are negligible in the Regge limit, and in this work, similar to our previous work~\cite{Xie:2019soz}, we use the gravitational form factor of the proton for $A(t)$, which was derived by the authors of Ref.~\cite{Abidin:2009hr} with the bottom-up AdS/QCD model.
We utilize the result obtained with the soft-wall model, in which the AdS geometry is smoothly cut off at the infrared boundary.
The resulting expression of the gravitational form factor includes some parameters, but those can be determined by the basic hadron properties, such as the proton mass.
Hence the form factor we adopt in this work does not bring any adjustable parameters.

The vertex of glueball-proton-proton in the Regge limit is given by 
\begin{equation}\label{vertex}
\Gamma_{g}^{\mu\nu}= \frac{i \lambda_{g}A(t)}{2}(\gamma^{\mu}P^{\nu} + \gamma^{\nu}P^{\mu}),
\end{equation}
where $\lambda_{g}$ is the coupling constant.
The massive spin-2 glueball propagator is written as \cite{Yamada:1982dx}
\begin{equation}\label{propagator}
D_{\alpha\beta\gamma\delta}^{g}(k) = \frac{-id_{\alpha\beta\gamma\delta}}{k^2 + m_{g}^2},
\end{equation}
where $m_{g}$ is the glueball mass, and $\alpha,\beta$ and $\gamma,\delta$ are Lorentz indices for the initial and final states.
$d_{\alpha\beta\gamma\delta}$ is explicitly expressed as
\begin{align}
   d_{\alpha\beta\gamma\delta} = &\frac{1}{2}(\eta_{\alpha\gamma}\eta_{\beta\delta} + \eta_{\alpha\delta}\eta_{\beta\gamma}) - \frac{1}{2m_{g}^2}(k_{\alpha}k_{\delta}\eta_{\beta\gamma} + k_{\alpha}k_{\gamma}\eta_{\beta\delta} + k_{\beta}k_{\delta}\eta_{\alpha\gamma} + k_{\beta}k_{\gamma}\eta_{\alpha\delta})\nonumber\\
                                &+\frac{1}{24}\left[\left(\frac{k^2}{m_{g}^2}\right)^2 - 3\left(\frac{k^2}{m_{g}^2}\right) - 6\right]\eta_{\alpha\beta}\eta_{\gamma\delta} - \frac{k^2 - 3m_{g}^2}{6m_{g}^4}(k_{\alpha}k_{\beta}\eta_{\gamma\delta} + k_{\gamma}k_{\delta}\eta_{\alpha\beta}) \nonumber\\
       &+\frac{2k_{\alpha}k_{\beta}k_{\gamma}k_{\delta}}{3m_{g}^4}.
\end{align}
With the exception of the first term, contributions of the others are negligible in the Regge limit.
Combining Eqs.~\eqref{vertex} and \eqref{propagator}, the glueball exchange amplitude in the $pp(p\bar{p})$ scattering is obtained as
\begin{align}\label{a_g}
\mathcal{A}_{g}^{pp(p\bar{p})} &= (\bar{u}_1\Gamma_{g}^{\alpha\beta}u_3)D_{\alpha\beta\gamma\delta}^{g}(k)(\bar{u}_2\Gamma_{g}^{\gamma\delta}u_4)\nonumber\\
   &= \frac{i\lambda_{g}^2}{8(k^2 + m_{g}^2)}[2sA^2(t)(\bar{u}_1\gamma^{\alpha}u_3)(\bar{u}_2\gamma_{\alpha}u_4)+4A^2(t)p_2^{\alpha}p_1^{\beta}(\bar{u}_1\gamma_{\alpha}u_3)(\bar{u}_2\gamma_{\beta}u_4)].
\end{align}

The vector meson propagator can be written as \cite{Anderson:2016zon}
\begin{equation}
D_{\mu\nu}^{v}(k) = \frac{i}{k^2 + m_{v}}\eta_{\mu\nu},
\end{equation}
where $m_{v}$ is the vector meson mass.
The vector-proton-proton vertex is written as 
 \begin{equation}
\Gamma_{v}^{\mu} = -i\lambda_{v}\gamma^{\mu},
\end{equation}
where $\lambda_{v}$ is the coupling constant.
These lead to the amplitude for the vector meson exchange in the $pp(p\bar{p})$ scattering 
\begin{align}\label{a_v}
\mathcal{A}_{v}^{pp(p\bar{p})}&=(\bar{u}_1\Gamma_{v}^{\mu}u_3)D_{\mu\nu}^{v}(k)(\bar{u}_2\Gamma_{v}^{\nu}u_4)\nonumber\\
&= -\frac{i\lambda_{v}^2}{k^2+m_{v}^2}\eta_{\mu\nu}(\bar{u}_1\gamma^{\mu}u_3)(\bar{u}_2\gamma^{\nu}u_4).
\end{align}

According to Eq.~\eqref{tot}, the total amplitude is obtained as
\begin{align}
    \mathcal{A}^{pp(p\bar{p})}_{\mathrm{tot}} =&\frac{-i\lambda_{g}^2}{8(t - m_{g}^2)}[2sA^2(t)(\bar{u}_1\gamma^{\alpha}u_3)(\bar{u}_2\gamma_{\alpha}u_4) + 4A^2(t)p_2^{\alpha}p_1^{\beta}(\bar{u}_1\gamma_{\alpha}u_3)(\bar{u}_2\gamma_{\beta}u_4)] \nonumber
     \\&+\frac{i\lambda_{v}^2}{t - m_{v}^2}\eta_{\mu\nu}(\bar{u}_1\gamma^{\mu}u_3)(\bar{u}_2\gamma^{\nu}u_4).
\end{align}
Taking into account the conditions, $s \gg |t|$ , $u_1\approx u_3$ and $u_2\approx u_4$, the differential cross section can be obtained as
\begin{align}\label{dcs1}
\frac{d\sigma}{dt} &= \frac{1}{16\pi s^2}|\mathcal{A}_{\mathrm{tot}}|^2\nonumber\\
                 &=\frac{\lambda_{g}^4s^2A^2(t)}{16\pi |t - m_{g}^2|^2} - \frac{\lambda_{g}^2\lambda_{v}^2A^2(t)s}{8\pi}\bigg[\frac{1}{(t-m_g^2)^*}\times\frac{1}{t-m_v^2}+\frac{1}{(t-m_v^2)^*}\times\frac{1}{t-m_g^2}\bigg]+\frac{\lambda_{v}^4}{4\pi|t-m_{v}^2|^2},
\end{align}
where the asterisk indicates complex conjugation.
The first, second and third terms of the right-hand side represent contributions of the glueball exchange, the cross term for both the glueball and vector meson exchanges, and the vector meson exchange, respectively.
Since in this equation only the lightest states are considered, the propagators need to be Reggeized to include the higher spin states on the Regge trajectories, which correspond to the excitation of strings in the curved spacetime.

We briefly review the Reggeization procedure, which was presented in detail in Ref.~\cite{Anderson:2016zon}.
The bosonic open string four-tachyon amplitude can be expressed as
\begin{equation}\label{a0}
\mathcal{A}_{o}^4(s,t,u) = \widetilde{\mathcal{A}}_{o}(s,t) + \widetilde{\mathcal{A}}_{o}(u,t) + \widetilde{\mathcal{A}}_{o}(s,u),
\end{equation}
where
\begin{equation}\label{os}
\widetilde{\mathcal{A}}_{o}(x,y) = iC\frac{\Gamma[-a_{o}(x)]\Gamma[-a_{o}(y)]}{\Gamma[-a_{o}(x)-a_{o}(y)]}.
\end{equation}
Here $a_{o}(x)=1+\alpha'_{o}x$, in which $\alpha'_{o}$ is the slope.
Since $s+t+u=4m_{p}^2$ and $\alpha_{v}(x)=\alpha_{v}(0)+\alpha'_{v}x$,
\begin{equation}\label{chi_v}
\alpha_{v}(s)+\alpha_{v}(t)+\alpha_{v}(u) = 3 + \alpha'_v ( 4 m_p^2 - 3 m_v^2) \equiv \chi_{v}.
\end{equation}
  Particles on the vector meson Regge trajectory have odd spins, and we need to replace the propagator of the vector meson with the exchange of odd spin string states. The $t$-channel is the dominant channel in the Regge limit, so we can ignore the contribution of $\widetilde{\mathcal{A}}_{o}(s,u)$, since it has no pole in the $t$-channel.
Then we replace $a_{o}(x)$ with $\alpha_{v}(x)$, and $u$ with a function of $s$ and $t$.
Combining Eqs.~\eqref{os} and \eqref{chi_v},
\begin{align}\label{a_u}
 &\widetilde{\mathcal{A}}_{o}(u,t)=iC\frac{\Gamma[\alpha_{v}(s)+\alpha_{v}(t)-\chi_{v}]\Gamma[-\alpha_{v}(t)]}{\Gamma[\alpha_{v}(s)-\chi_{v}]},
 \\&\widetilde{\mathcal{A}}_{o}(s,t)=iC\frac{\Gamma[-\alpha_{v}(s)]\Gamma[-\alpha_{v}(t)]}{\Gamma[-\alpha_{v}(s)-\alpha_{v}(t)]}.
\end{align}
In order to obtain the odd spin states, we need to take the difference between $\widetilde{\mathcal{A}}_{o}(u,t)$ and $\widetilde{\mathcal{A}}_{o}(s,t)$. Therefore the amplitude of the Reggeon exchange can be rewritten as 

\begin{equation}
\mathcal{A}_{R}^4=iC\frac{\Gamma[\alpha_{v}(s)+\alpha_{v}(t)-\chi_{v}]}{\Gamma[\alpha_{v}(s)-\chi_{v}]}-iC\frac{\Gamma[-\alpha_{v}(s)]\Gamma[-\alpha_{v}(t)]}{\Gamma[-\alpha_{v}(s)-\alpha_{v}(t)]}.
\end{equation}
It is found that the amplitude is asymmetric under exchanges of $s$, $t$ and $u$, and the $pp$ and $p \bar p$ scattering processes have different behavior in the Reggeon exchange. This behavior is consistent with the experimental data, and $\lambda_{v}$ values of vector-proton-proton and vector-antiproton-antiproton cases are different from each other.
$\mathcal{A}_{R}^4$ can be expanded around the $\alpha_{v}(t)=1$ pole in the Regge limit as
\begin{equation}
\mathcal{A}_{R}^4\approx iC\big(1-e^{-i\pi \alpha_{v}(t)}\big)\big(\alpha'_{v}s\big)^{\alpha_{v}(t)}\Gamma[-\alpha_{v}(t)].
\end{equation}
Comparing this equation to Eq.~\eqref{a_v} and the third term of the right-hand side of Eq.~\eqref{dcs1}, it is found that the propagator of the vector meson needs to be replaced with
\begin{equation}
\frac{1}{t-m_{v}^2}  \rightarrow  \alpha'_{v}e^{-\frac{i\pi \alpha_{v}(t)}{2}}\sin\left[\frac{\pi\alpha_{v}(t)}{2}\right]\left(\alpha'_{v}s\right)^{\alpha_{v}(t)-1}\Gamma[-\alpha_{v}(t)].
\end{equation}

The bosonic closed string four-tachyon amplitude is given by
\begin{equation}
\mathcal{A}_{c}^4(s,t,u)=2\pi C\frac{\Gamma\big[-\frac{a_{c}(t)}{2}\big]\Gamma\big[-\frac{a_{c}(s)}{2}\big]\Gamma\big[-\frac{a_{c}(u)}{2}\big]}{\Gamma\big[-\frac{a_{c}(s)}{2}-\frac{a_{c}(t)}{2}\big]\Gamma\big[-\frac{a_{c}(s)}{2}-\frac{a_{c}(u)}{2}\big]\Gamma\big[-\frac{a_{c}(t)}{2}-\frac{a_{c}(u)}{2}\big]},
\end{equation}
where $a_{c}(x)=2+a'_{c}x/2$.
Repeating the operation for Eq.~\eqref{a_u}, one can obtain 
\begin{equation}
\mathcal{A}_{c}^4(s,t)=2\pi C\frac{\Gamma\big[-\frac{a_{c}(t)}{2}\big]\Gamma\big[-\frac{a_{c}(s)}{2}\big]\Gamma\big[1+\frac{a_{c}(s)}{2}+\frac{a_{c}(t)}{2}\big]}{\Gamma\big[-\frac{a_{c}(s)}{2}-\frac{a_{c}(t)}{2}\big]\Gamma\big[-\frac{a_{c}(s)}{2}-\frac{a_{c}(u)}{2}\big]\Gamma\big[-\frac{a_{c}(t)}{2}-\frac{a_{c}(u)}{2}\big]}.
\end{equation}
The bosonic closed string states only include even spin particles, and we need to replace $a_{c}(t)+2$ with $\alpha_{g}(t)$, due to the fact that the first pole corresponds to the spin-2 glueball. Then the amplitude of the Pomeron exchange can be expressed as
\begin{equation}
\mathcal{A}_{P}^4(s,t) = 2\pi C\frac{\Gamma\big[1-\frac{\alpha_{g}(t)}{2}\big]\Gamma\big[1-\frac{\alpha_{g}(s)}{2}\big]\Gamma\big[1-\frac{\chi_{g}}{2}+\frac{\alpha_{g}(s)}{2}+\frac{\alpha_{g}(t)}{2}]}
{\Gamma\big[2-\frac{\alpha_{g}(t)}{2}-\frac{\alpha_{g}(s)}{2}\big]\Gamma\big[2-\frac{\chi_{g}}{2}+\frac{\alpha_{g}(s)}{2}\big]\Gamma\big[2-\frac{\chi_{g}}{2}+\frac{\alpha_{g}(t)}{2}\big]},
\end{equation}
where $\chi_{g} \equiv \alpha_{g}(s)+\alpha_{g}(u)+\alpha_{g}(t)$, in which $\alpha_{g}(x)=\alpha_{g}(0)+\alpha'_{g}x$. Expanding the above equation around the $\alpha_{g}(t) = 2$ pole in the Regge limit, one obtains
\begin{equation}
\mathcal{A}_{P}^4\approx 2\pi Ce^{-\frac{i\pi\alpha_{g}(t)}{2}}\left(\frac{\alpha'_{g}s}{2}\right)^{\alpha_{g}(t)-2}\frac{\Gamma\left[1-\frac{\alpha_{g}(t)}{2}\right]}{\Gamma\left[2-\frac{\chi_{g}}{2}+\frac{\alpha_{g}(t)}{2}\right]}.
\end{equation}
Comparing this equation to Eq.~\eqref{a_g} and the first term of the right-hand side of Eq.~\eqref{dcs1}, it is found that the propagator of massive spin-2 glueball needs to be replaced with
\begin{equation}
\frac{1}{t-m_{g}^2} \rightarrow  \frac{\alpha'_{g}}{2}e^{-\frac{i\pi\alpha_{g}(t)}{2}}\left(\frac{\alpha'_{g}s}{2}\right)^{\alpha_{g}(t)-2}\frac{\Gamma\left[3-\frac{\chi_{g}}{2}\right]\Gamma\left[1-\frac{\alpha_{g}(t)}{2}\right]}{\Gamma\left[2-\frac{\chi_{g}}{2}+\frac{\alpha_{g}(t)}{2}\right]}.
\end{equation}

With the Reggeized propagators introduced above, the differential cross section can be obtained as
\begin{align}
 \frac{d\sigma}{dt}=&\frac{\lambda_{g}^4s^2A^4(t)}{16\pi}\left[ \frac{\alpha'_{g}}{2}\frac{\Gamma\left[3-\frac{\chi_{g}}{2}\right]\Gamma\left[1-\frac{\alpha_{g}(t)}{2}\right]}{\Gamma\left[2-\frac{\chi_{g}}{2}+\frac{\alpha_{g}(t)}{2}\right]}\left(\frac{\alpha'_{g}s}{2}\right)^{\alpha_{g}(t)-2}\right]^2\nonumber
\\&-\frac{\lambda_{g}^2\lambda_{v}^2sA^2(t)}{4\pi}\left[ \frac{\alpha'_{g}}{2}\frac{\Gamma\left[3-\frac{\chi_{g}}{2}\right]\Gamma\left[1-\frac{\alpha_{g}(t)}{2}\right]}{\Gamma\left[2-\frac{\chi_{g}}{2}+\frac{\alpha_{g}(t)}{2}\right]}\left(\frac{\alpha'_{g}s}{2}\right)^{\alpha_{g}(t)-2}\right]\nonumber
\\&\qquad\times\left[\alpha'_{v}\sin\bigg(\frac{\pi\alpha_{v}(t)}{2}\bigg)(\alpha'_{v}s)^{\alpha_{v}(t)-1}\Gamma[-\alpha_{v}(t)]\right]\cos \left[ \frac{\pi}{2}(\alpha_g(t)-\alpha_v(t)) \right] \nonumber
\\&+\frac{\lambda_{v}^4}{4\pi}\left[\alpha'_{v}\sin\bigg(\frac{\pi\alpha_{v}(t)}{2}\bigg)(\alpha'_{v}s)^{\alpha_{v}(t)-1}\Gamma[-\alpha_{v}(t)]\right]^2,
\label{eq_dcs_final}
\end{align}
which leads to the invariant amplitude for the Pomeron and Reggeon exchanges:
\begin{align}
\mathcal{A}(s,t) = &-s\lambda_{g}^2A^2(t)e^{-\frac{i\pi\alpha_{g}(t)}{2}}\frac{\Gamma\left[3-\frac{\chi_{g}}{2}\right]\Gamma\left[1-\frac{\alpha_{g}(t)}{2}\right]}{\Gamma\left[2-\frac{\chi_{g}}{2}+\frac{\alpha_{g}(t)}{2}\right]}\left(\frac{\alpha'_{g}s}{2}\right)^{\alpha_{g}(t)-1}\nonumber\\
&+2s\lambda_{v}^2\alpha'_{v}e^{-\frac{i\pi\alpha_{v}(t)}{2}}\sin\left(\frac{\pi\alpha_{v}(t)}{2}\right)(\alpha'_{v}s)^{\alpha_{v}(t)-1}\Gamma[-\alpha_{v}(t)].
\end{align}
The first and second terms of the right-hand side represent contributions of the Pomeron and Reggeon exchanges, respectively.
Applying the optical theorem, the total cross section is obtained as
\begin{align}
\sigma_{tot} = &\frac{1}{s} \mathrm{Im} \mathcal{A}(s,t=0)\nonumber\\
=&\lambda_{g}^2\sin\left(\frac{\pi\alpha_{g}(0)}{2}\right)\frac{\Gamma\left[3-\frac{\chi_{g}}{2}\right]\Gamma\left[1-\frac{\alpha_{g}(t)}{2}\right]}{\Gamma\left[2-\frac{\chi_{g}}{2}+\frac{\alpha_{g}(t)}{2}\right]}\left(\frac{\alpha'_{g}s}{2}\right)^{\alpha_{g}(0)-1}\nonumber\\
&-2\lambda_{v}^2\alpha'_{v}\sin^2 \left( \frac{\pi\alpha_{v}(0)}{2} \right) (\alpha'_{v}s)^{\alpha_{v}(0)-1}\Gamma[-\alpha_{v}(0)].
\label{eq_tcs_final}
\end{align}

\subsection{Contribution ratios of the Pomeron and Reggeon exchanges}
\label{sec:subsec_ratios}
\begin{figure}[t]
\begin{tabular}{cc}
\includegraphics[width=0.45\textwidth]{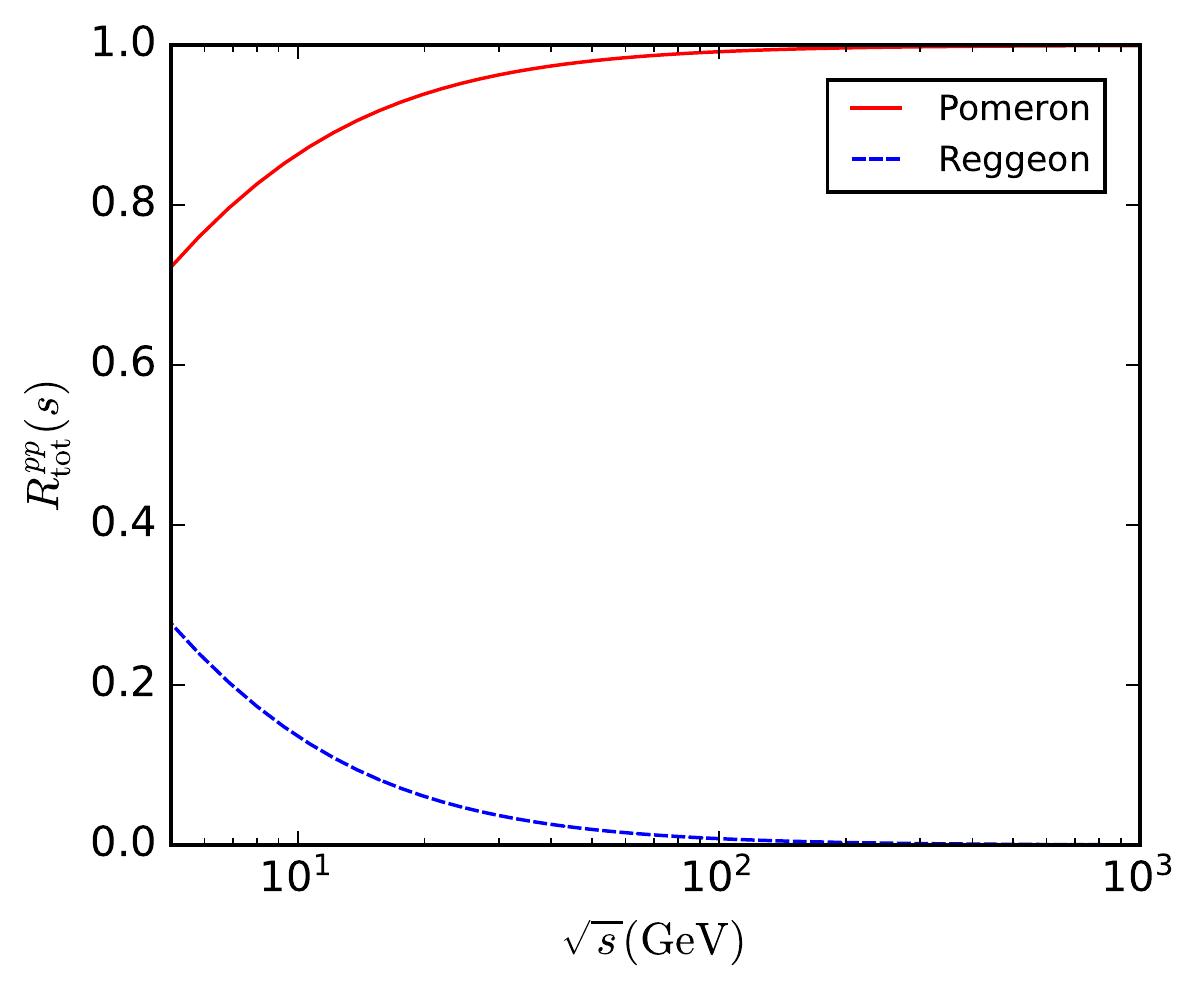}
\includegraphics[width=0.45\textwidth]{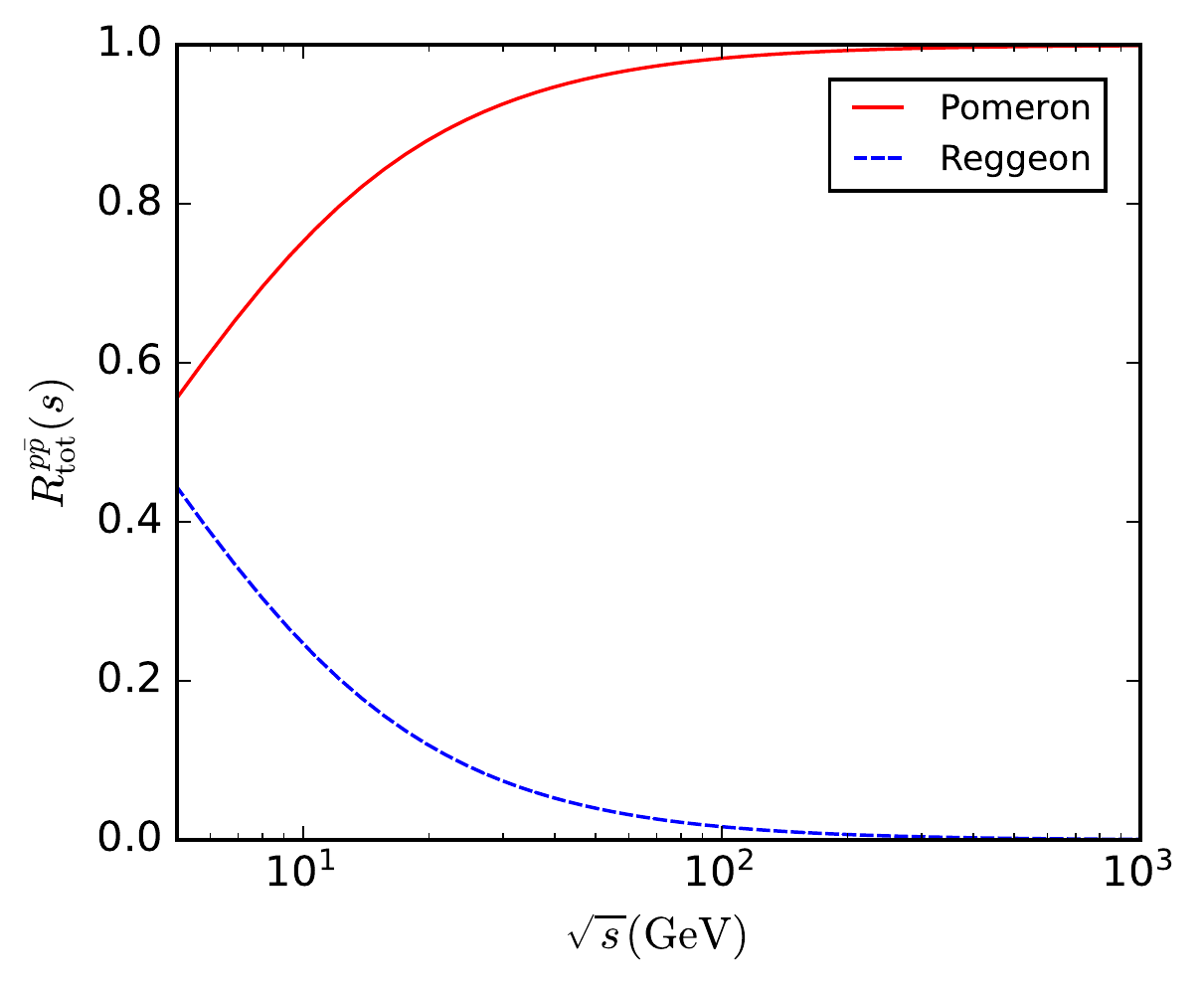}
\end{tabular}
\caption{
The contribution ratios for the total cross section as a function of $\sqrt{s}$.
The left and right panels are for the $pp$ and $p\bar{p}$ scattering, respectively.
The solid and dashed curves represent the results for the Pomeron and Reggeon exchanges, respectively.
}
\label{fig2}
\end{figure}

Here we present the energy dependence of contributions of the Pomeron and Reggeon exchanges in the present model.
The parameters we used for this analysis are the best fit values which will be explained in detail in the next section.
We numerically evaluate the Pomeron and Reggeon exchange contributions to the total cross sections and divide by the overall magnitude separately.
We define these as the contribution ratios, $R_{\mathrm{tot}}^{pp}$ and $R_{\mathrm{tot}}^{p\bar{p}}$, for the $pp$ and $p\bar{p}$ total cross sections, respectively.
Focusing on the kinematic range of $5 < \sqrt{s} < 1000$~GeV, we display the energy dependence of the ratios in Fig.~\ref{fig2}.
It can be seen from the left panel, which shows the $pp$ case, that the Pomeron contribution increases with the energy, and it is opposite for the Reggeon contribution.
At $\sqrt{s} \gtrsim 100$~GeV, the Reggeon contribution almost vanishes and the Pomeron contribution becomes dominant.
The results for the $p\bar{p}$ case, which are shown in the right panel, are similar, but the Reggeon contribution is somewhat larger compared to the $pp$ case in all the considered kinematic region.
From these results, we find that it is important to take into account the contribution of the Reggeon exchange, unless the energy is high enough.

\begin{figure}[t]
\begin{tabular}{cc}
\includegraphics[width=0.45\textwidth]{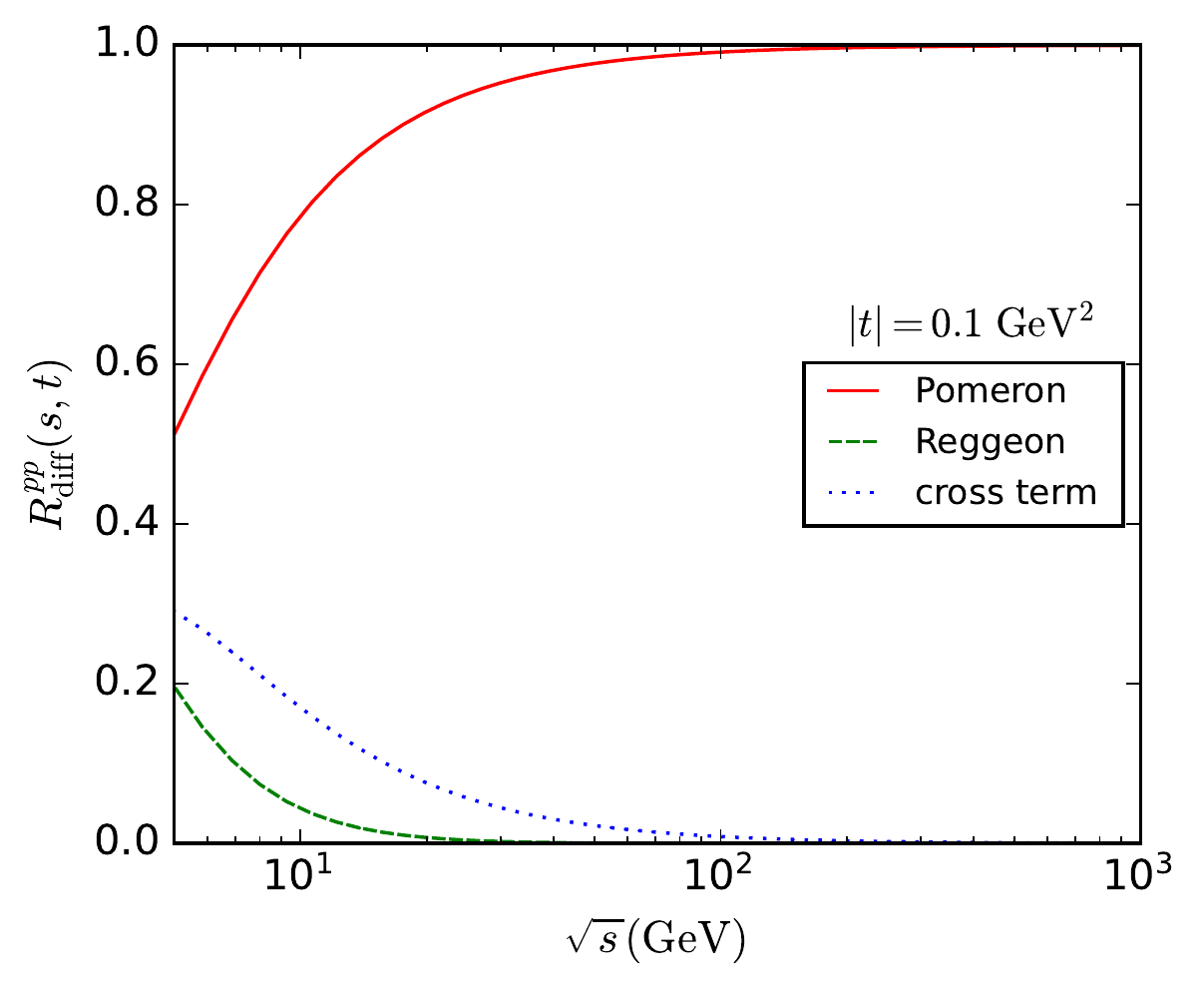}
\includegraphics[width=0.45\textwidth]{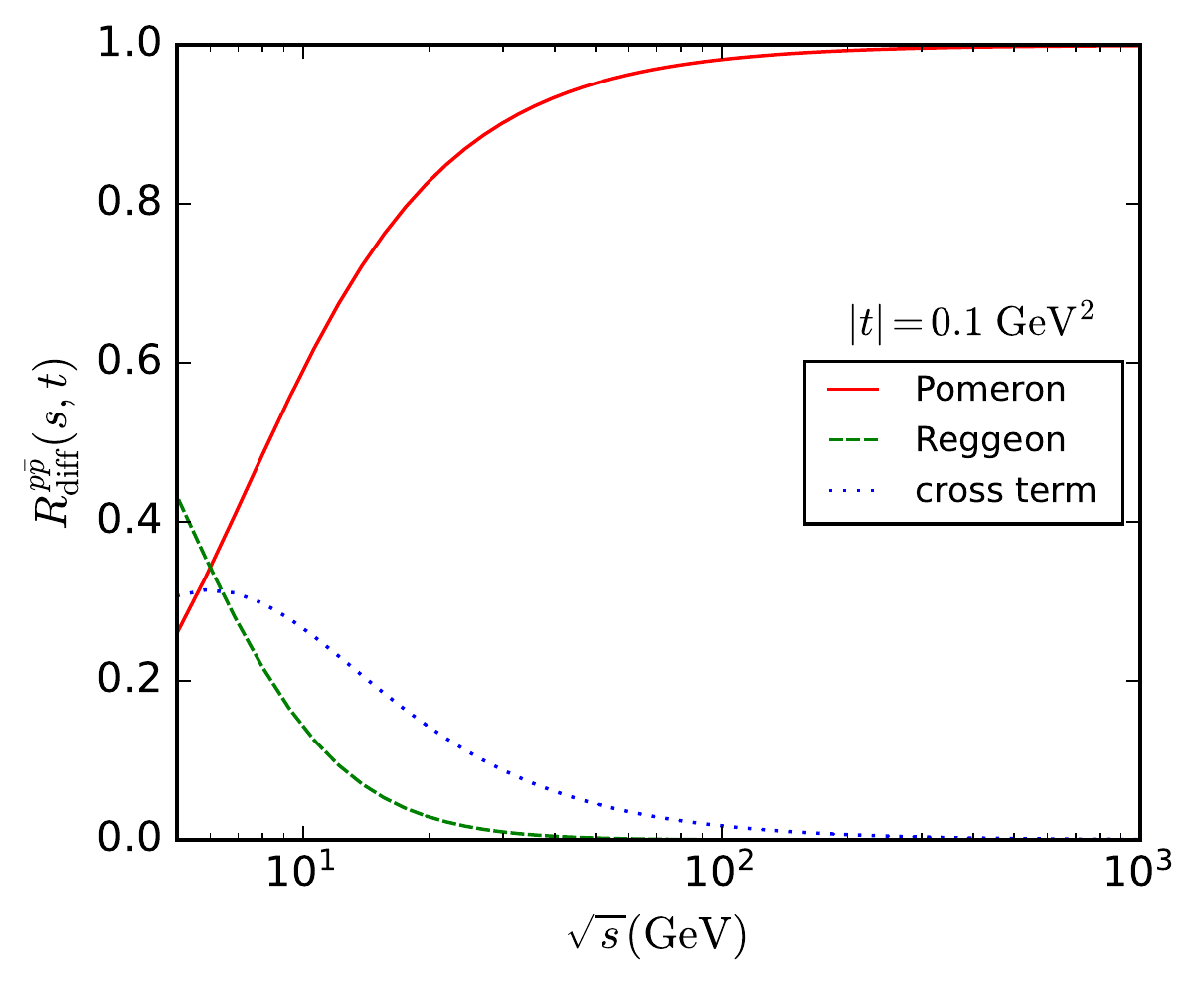}
\end{tabular}
\caption{
The contribution ratios for the differential cross section at $|t|=0.1$~GeV$^2$ as a function of $\sqrt{s}$.
The left and right panels are for the $pp$ and $p\bar{p}$ scattering, respectively.
The solid, dashed and dotted curves represent results for the Pomeron exchange, the Reggeon exchange and the cross term, respectively.
}
\label{fig3}
\end{figure}

Then we present results of the similar evaluations for the differential cross section, defining the contribution ratios, $R_{\mathrm{diff}}^{pp}$ and $R_{\mathrm{diff}}^{p\bar{p}}$, for the $pp$ and $p\bar{p}$ scattering, respectively.
The results obtained at $|t| = 0.1$~GeV$^2$ are shown in Fig.~\ref{fig3}.
Differently from the total cross section case, the resulting ratios for the cross term, which corresponds to the second term of the right-hand side of Eq.~\eqref{eq_dcs_final}, are also displayed.
The overall behavior of the ratios for the Pomeron and Reggeon exchanges is similar to the total cross section case.
The cross term contributions for both the $pp$ and $p\bar{p}$ cases decrease with the energy.
Due to the presence of the cross term, we find that it is more important to consider the Reggeon exchange contribution when we investigate the differential cross section, compared to the total cross section case.

\section{Numerical results}
\label{sec:results}

\subsection{Fitting procedure}
\label{sec:subsec_fitting_procedure}
The expressions for the cross sections presented in the previous section include seven adjustable parameters in total.
Three of them are associated with the Pomeron exchange, i.e., the intercept, slope and proton-glueball coupling constant.
Since for the Reggeon exchange the coupling constant ($\lambda_{v}$) depends on the scattering process, we need two ($\lambda_{vpp}$ and $\lambda_{v\bar{p}\bar{p}}$) for the $pp$ and $p\bar{p}$ cases, which is the reason why there are four parameters for the Reggeon exchange.
As to the former three parameters, we use the values obtained in the previous work~\cite{Xie:2019soz}, in which only the Pomeron exchange was considered to study the cross sections at $\sqrt{s} \geq 546$~GeV.
Since the Reggeon exchange contribution almost completely vanishes in such a high energy region, which can be seen from Figs.~\ref{fig2} and \ref{fig3}, this is justified.
The values, $\alpha_{g}(0)=1.084$, $\alpha'_{g}=0.368~\mathrm{GeV^{-2}}$ and $\lambda_{g}=9.59~\mathrm{GeV^{-1}}$, are used in the present analysis.
There are four parameters yet to be determined, but we can reduce to three by utilizing the linear relation for the Reggeon trajectory, $J=\alpha_{v}(0)+m_{v}^2\alpha'_{v}$, regarding the $\rho$ meson as the exchanged Reggeon with its physical mass.
Hence we find that it is enough to determine only the three parameters, $\alpha_{v}(0)$, $\lambda_{vpp}$ and $\lambda_{v\bar{p}\bar{p}}$, with the experimental data.

In this work we determine those three parameters, using the analytical result, Eq.~\eqref{eq_tcs_final}, and the experimental data of the total cross section.
All the currently available data, which are summarized by the Particle Data Group
(PDG) in 2020~\cite{ParticleDataGroup:2020ssz}, in the range of $5 \leq\sqrt{s}\leq1000~\mathrm{GeV}$ are taken into account.
For the numerical fitting we utilize the MINUIT package~\cite{James:1975dr}.
Once all the three parameters are determined, we can predict the differential cross section for both the $pp$ and $p\bar{p}$ scattering without any additional parameters.

\subsection{Fitting results}
\label{sec:subsec_fitting_results}
\begin{figure}[t]
\begin{center}
\includegraphics[width=0.75\textwidth]{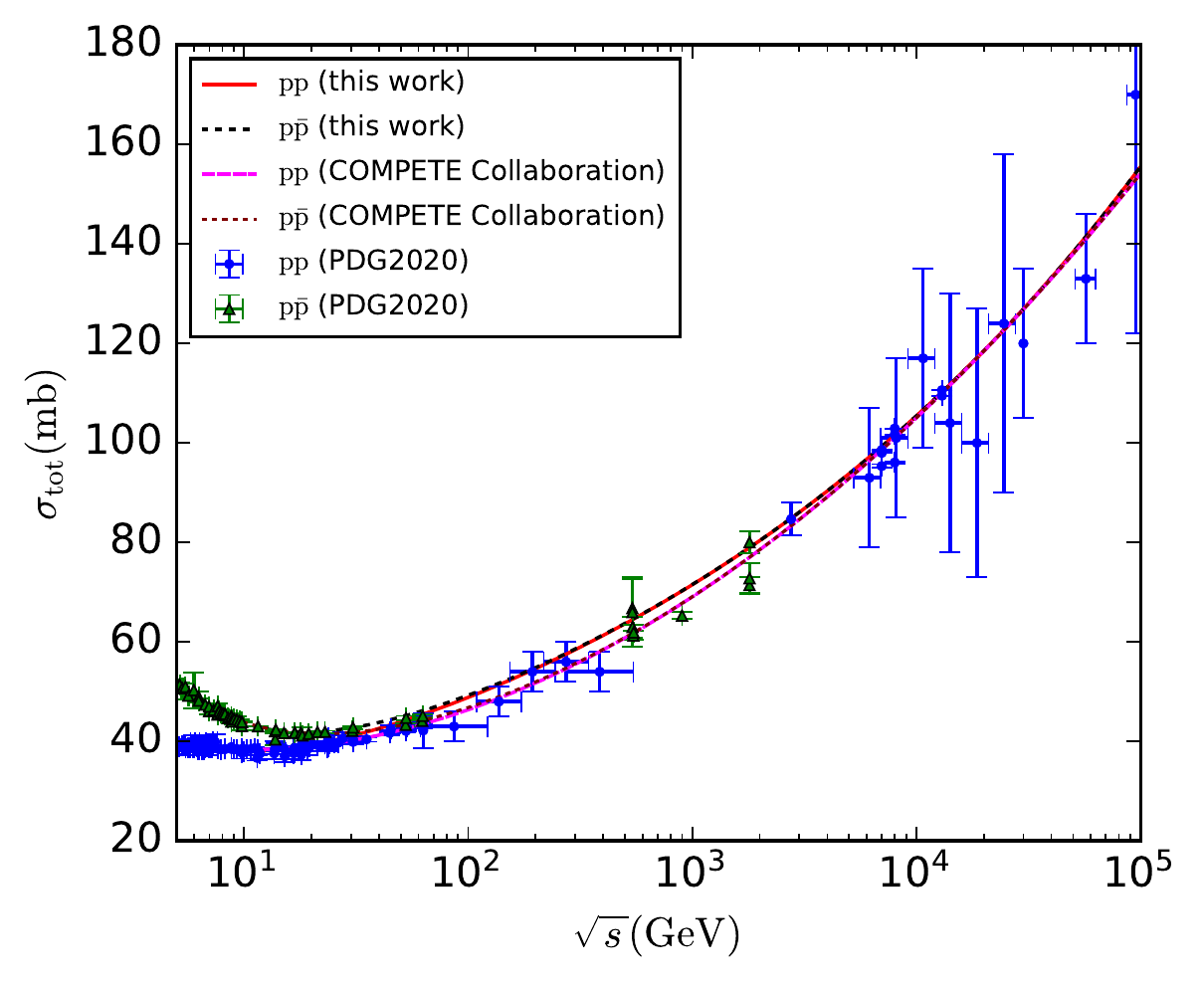}
\caption{\label{fig4}
The total cross section of the $pp$ and $p\bar{p}$ scattering as a function of $\sqrt{s}$.
Our calculations are compared with the results obtained by the COMPETE Collaboration~\cite{COMPETE:2002jcr}.
The experimental data are taken from Ref.~\cite{ParticleDataGroup:2020ssz}.
}
\end{center}
\end{figure}

The resulting best fit values of the three parameters are found to
be: $\alpha_{v}(0)=0.444 \pm 0.008$, $\lambda_{vpp}=7.742 \pm 0.205$ and $\lambda_{v\bar{p}\bar{p}}=16.127 \pm 0.295$.
Combining both the Pomeron and Reggeon exchange contributions, we display our calculations of the total cross section for the $pp$ and $p\bar{p}$ scattering in Fig.~\ref{fig4},
in which the results obtained by the COMPETE Collaboration~\cite{COMPETE:2002jcr} are also shown for comparison.
It is found that our results are in agreement with the data in all the considered kinematic region for both the $pp$ and $p\bar{p}$ scattering.
Ours are also consistent with the results obtained by the COMPETE Collaboration.

\begin{figure}[t]
\begin{center}
\includegraphics[width=0.93\textwidth]{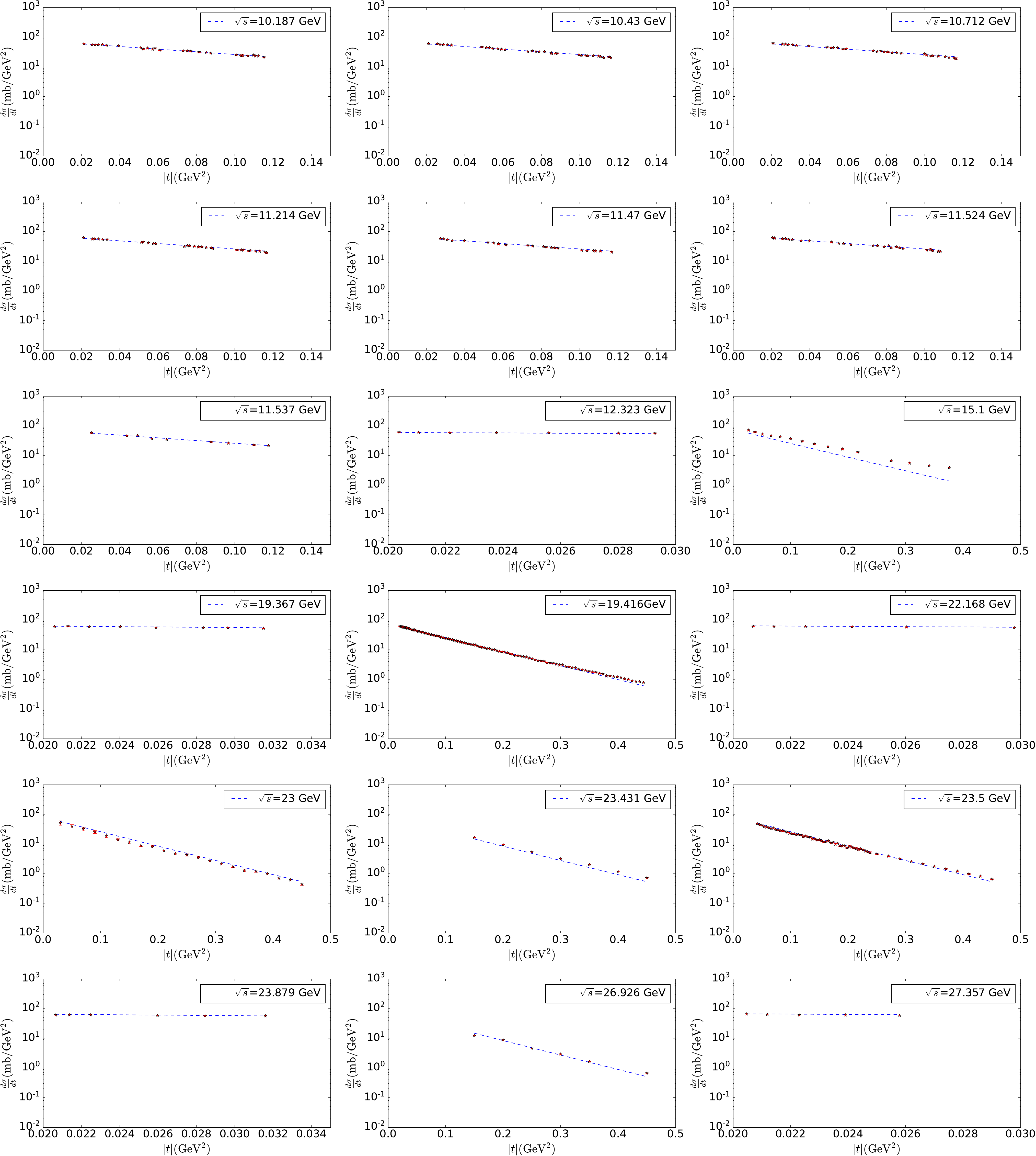}
\caption{\label{fig5}
The differential cross section of the $pp$ scattering as a function of $|t|$ for $10 < \sqrt{s} < 30~\mathrm{GeV}$.
The dashed curves represent our calculations, and the experimental data are depicted by stars with error bars.
}
\end{center}
\end{figure}

\begin{figure}[t]
\begin{center}
\includegraphics[width=0.93\textwidth]{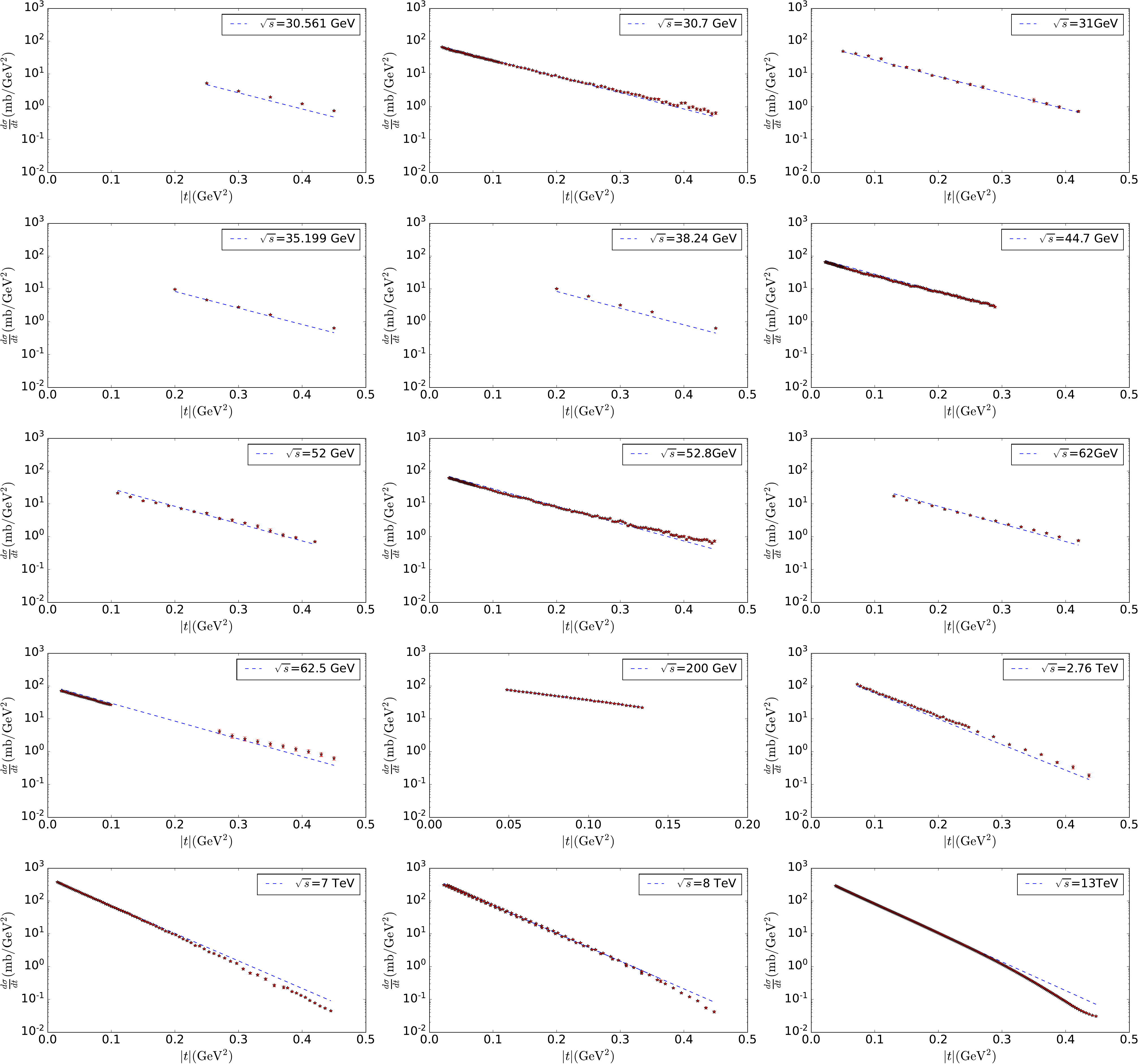}
\caption{\label{fig6}
The differential cross section of the $pp$ scattering as a function of $|t|$ for 30~GeV $< \sqrt{s} \leq 13$~TeV.
The dashed curves represent our calculations, and the experimental data are depicted by stars with error bars.
}
\end{center}
\end{figure}

\begin{figure}[t]
\begin{center}
\includegraphics[width=0.93\textwidth]{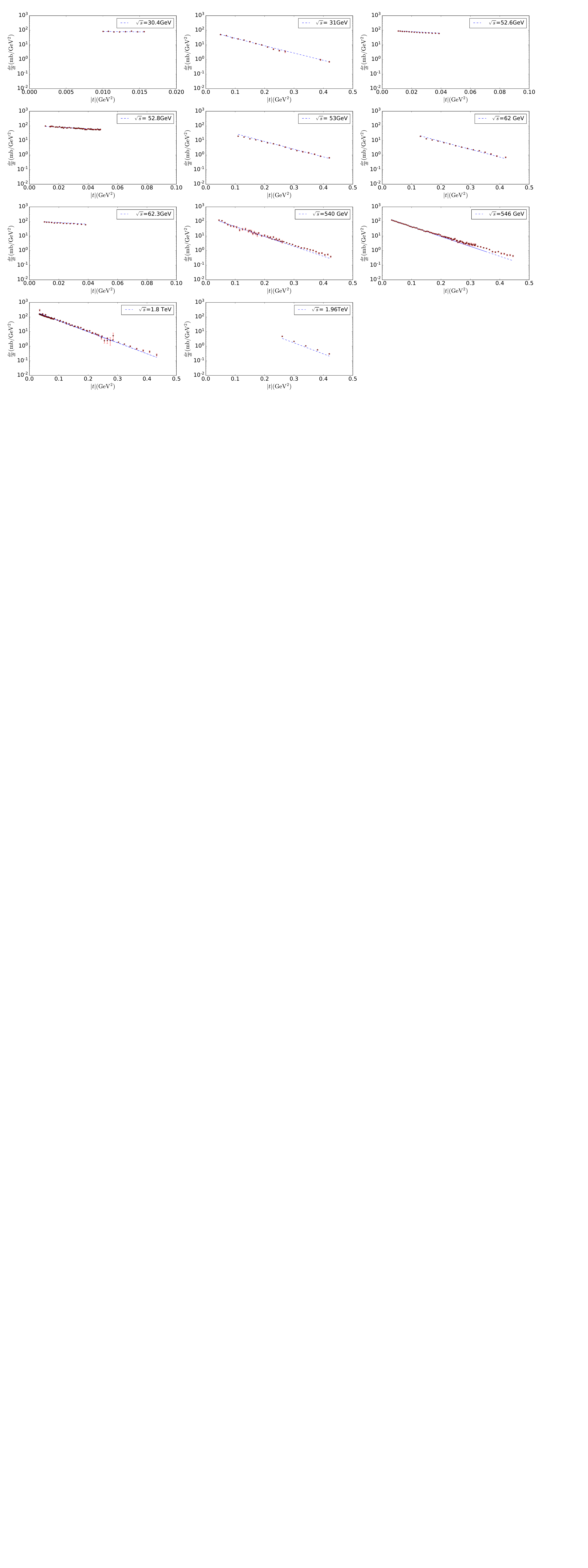}
\caption{\label{fig7}
The differential cross section of the $p\bar{p}$ scattering as a function of $|t|$ for $\sqrt{s} > 30$~GeV.
The dashed curves represent our calculations, and the experimental data are depicted by stars with error bars.
}
\end{center}
\end{figure}

Then we present our predictions for the differential cross section, focusing on the Regge regime.
Since we cannot neglect the Coulomb interaction in the very small $|t|$ region~\cite{Amos:1985wx} and also we need to exclude the diffractive minimum, the kinematic range of $0.01 < |t| < 0.45$~GeV$^2$ and $10$~GeV $< \sqrt{s} \leq 13$~TeV is considered.
The experimental data, to which we compare our calculations, are taken from Refs.~\cite{CHLM:1976vgz,AMES-BOLOGNA-CERN-DORTMUND-HEIDELBERG-WARSAW:1984yby,TOTEM:2013lle,Amaldi:1979kd,TOTEM:2016lxj,Kwak:1975yq,TOTEM:2011vxg,STAR:2020phn,TOTEM:2018psk,TOTEM:2018hki,Fidecaro:1981dk,Beznogikh:1973um,Jenkins:1978fv,Geshkov:1976ku,Schiz:1979rh} and Refs.~\cite{CERN-Naples-Pisa-StonyBrook:1982lzk,CDF:1993qdf,UA1:1982nds,E-710:1990vqb,UA4:1983mlb,D0:2012erd,AMES-BOLOGNA-CERN-DORTMUND-HEIDELBERG-WARSAW:1984yby} for the $pp$ and $p\bar{p}$ scattering, respectively.
For display purpose we split the results for the $pp$ scattering into two figures.
The $pp$ results for the kinematic range of $10 < \sqrt{s} < 30~\mathrm{GeV}$ are shown in Fig.~\ref{fig5}.
Except for the result at $\sqrt{s} = 15.1$~GeV, it can be seen that overall our calculations are consistent with the data.
The results for 30~GeV $< \sqrt{s} \leq 13$~TeV are displayed in Fig.~\ref{fig6}.
A quite wide $\sqrt{s}$ range is covered in this figure, but it is found that our model well describes the data.
Even so, some deviations at $|t| \sim 0.4$~GeV$^2$ can be seen especially for the results in the TeV scale, which may imply the applicable limit of the present model.
The results for the $p\bar{p}$ scattering are presented in Fig.~\ref{fig7}.
Similar to the $pp$ case, it is found that overall our calculations are consistent with the data in the wide $\sqrt{s}$ range.
Some deviations at $|t| \sim 0.4$~GeV$^2$ can be seen for the result at $\sqrt{s} = 546$~GeV, but the higher $\sqrt{s}$ results seem better.

\begin{figure}[t]
\begin{tabular}{cc}
\includegraphics[width=0.45\textwidth]{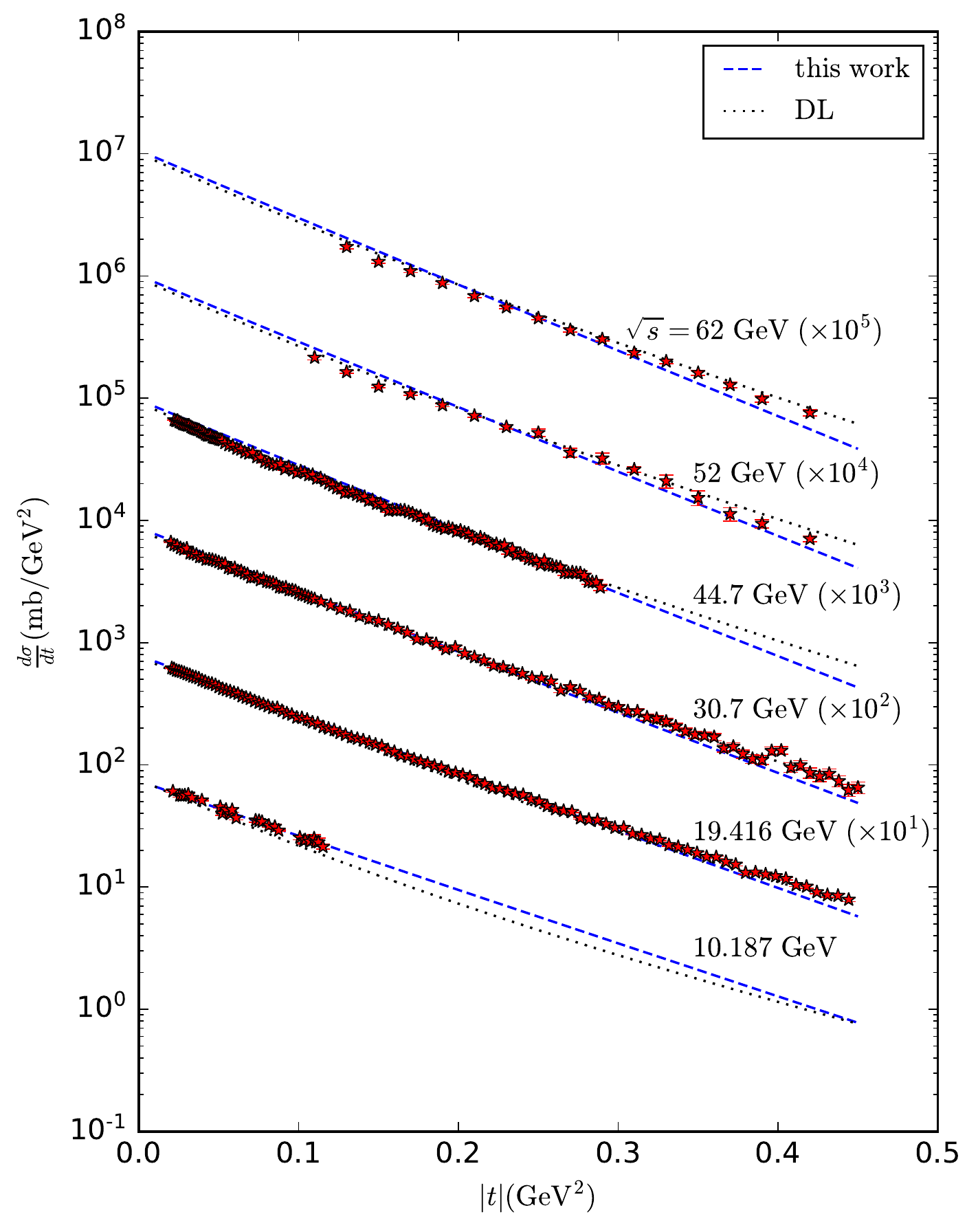}
\includegraphics[width=0.45\textwidth]{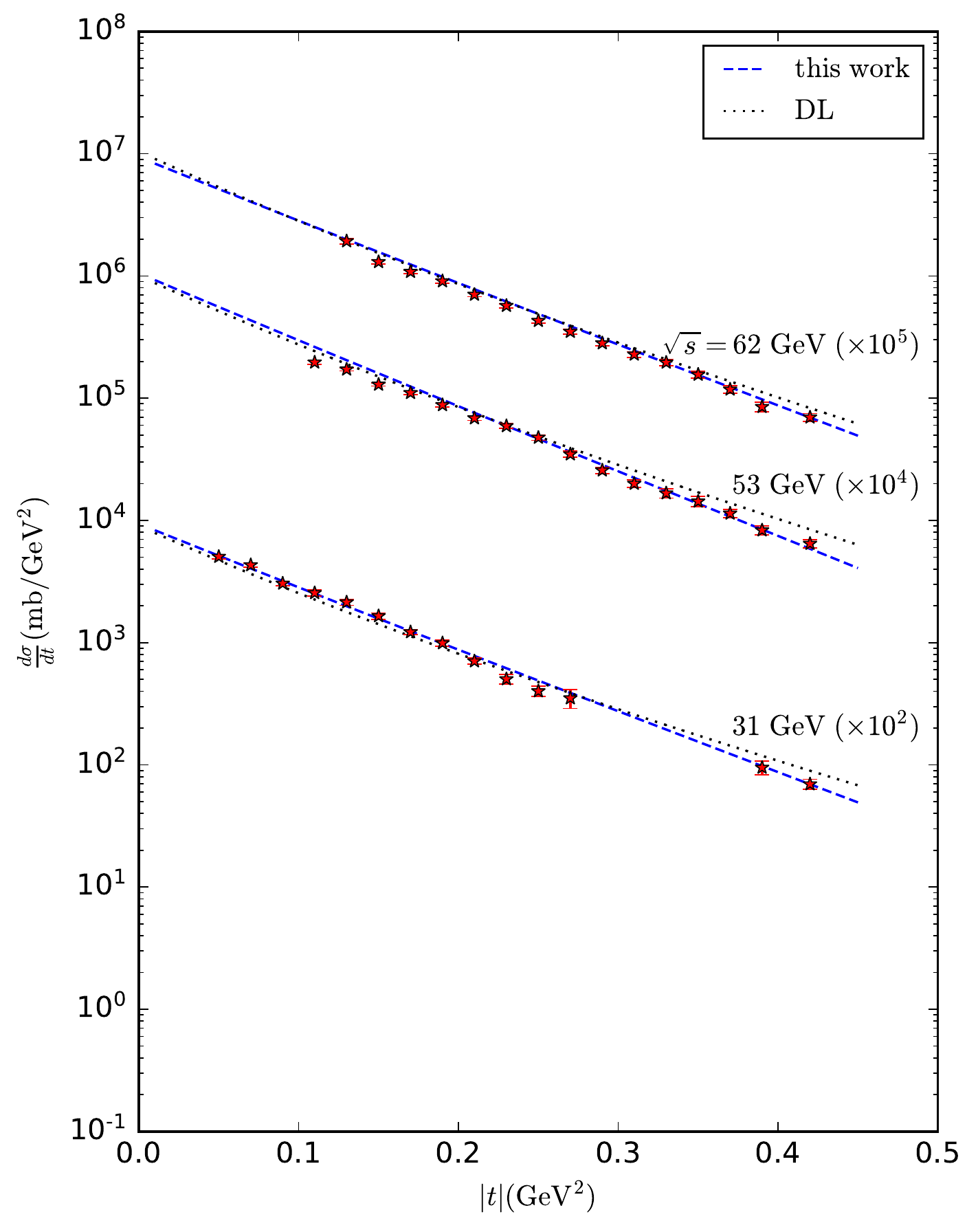}
\end{tabular}
\caption{
The differential cross section of the $pp$ (left) and $p\bar{p}$ (right) scattering as a function of $|t|$ for various $\sqrt{s}$.
The dashed and dotted curves represent our calculations and the results obtained by Donnachie and
Landshoff~\cite{Donnachie:1983hf}, respectively.
}
\label{fig8}
\end{figure}

Finally, for several values of $\sqrt{s}$ we present in Fig.~\ref{fig8}
comparisons for the $pp$ and $p\bar{p}$ differential cross sections between our predictions and the results obtained by Donnachie and
Landshoff~\cite{Donnachie:1983hf}, in which both the Pomeron and Reggeon contributions are taken into account.
It can be seen that our results are consistent with theirs in the considered kinematic region.
As to the Reggeon trajectory, the resulting intercept and slope of this work are quite close to those in their work.
However, our slope of the Pomeron trajectory is somewhat different from theirs, which is reflected in the slightly different $t$ dependence between our calculations and their results.

\section{Conclusion}
\label{sec:conclusion}
We have investigated the elastic $pp$ and $p\bar{p}$ scattering in a holographic QCD model, taking into account both the Pomeron and Reggeon exchanges in the Regge regime.
In our model setup, the Pomeron and Reggeon exchanges are described by the Reggeized spin-2 glueball and vector meson propagators, respectively.
Combining the proton-vector meson and proton-glueball couplings with those propagators, the scattering amplitudes are obtained.
We have explicitly presented the energy dependence of the Pomeron and Reggeon exchange contributions, focusing on the contribution ratios, from which we have found that it is important to consider the Reggeon exchange contribution, unless the energy is high enough.

There are several parameters in the model, but the ones associated with the Pomeron exchange have already been determined in the previous work~\cite{Xie:2019soz}.
Hence we have performed the numerical evaluations with only three adjustable parameters, which are determined with the experimental data of the total cross section for the $pp$ and $p \bar p$ scattering.
We have shown that the resulting total cross sections are in agreement with the data in a wide kinematic region.
Then it has been presented that our predictions for the differential cross sections are also consistent with the data in the forward region for a wide $\sqrt{s}$ range.

Through this study, we have found that we can well describe the forward $pp$ and $p \bar p$ scattering in the much wider kinematic region, compared to the previous work.
Since the results presented in this paper imply that the present model may have the strong predictive power for the various hadron-hadron forward scattering processes, further studies are certainly needed.
Also, future directions include applications to other high energy scattering processes, such as deep inelastic scattering or photoproduction of vector mesons.
To realize these, it is obvious that the conventional soft Pomeron with the constant intercept is not enough.
However, recently the authors of Ref.~\cite{Dosch:2022mop} have shown that the soft to hard Pomeron transition can be described with a scale-dependent Pomeron obtained in holographic light-front QCD.
It is important to describe the energy scale dependence of the Pomeron intercept within the present model, which will make further applications possible.
Finally, more experimental data are necessary to better constrain the model.
It is expected that future experiments of high energy hadron scattering will help to deepen our understandings of the nonperturbative nature of the strong interaction.

\section*{Acknowledgments}
The work of W.X. was supported by the National Natural Science Foundation of China under Grant No. 11875178.
The work of A.W. was supported by the start-up funding from China Three Gorges University.
A.W. is also grateful for the support from the Chutian Scholar Program of Hubei Province.

\bibliography{references}

\end{document}